# GaN/AlGaN microcavities for enhancement of non linear optical effects

V. Tasco[a], I. Tarantini[a], A. Campa[a], A. Massaro[a], T. Stomeo[a], G. Epifani[a], A. Passaseo[a], M. Braccini[b], M.C. Larciprete[b], C. Sibilia[b], F.A. Bovino[c]

[a] National Nanotechnology Laboratory of CNR-INFM, Distretto Tecnologico-ISUFI, Università del Salento, Via Arnesano, 73100 Lecce, Italy;
[b]Dipartimento di Energetica- Università di Roma "La Sapienza", via A. Scarpa 16, 00161 Roma, Italy;
[c] Quantum Optics Laboratory, Selex-SI, Via Puccini 2 Genova, Italy

## ABSTRACT

We present a study on the design, growth and optical characterization of a GaN/AlGaN microcavity for the enhancement of second order non linear effects. The proposed system exploits the high second order nonlinear optical response of GaN due to the non centrosymmetric crystalline structure of this material. It consists of a GaN cavity embedded between two GaN/AlGaN Distributed Bragg Reflectors designed for a reference mode coincident with a second harmonic field generated in the near UV region (~ 400 nm). Critical issues for this target are the crystalline quality of the material, together with sharp and abrupt interfaces among the multi-stacked layers. A detailed investigation on the growth evolution of GaN and AlGaN epilayers in such a configuration is reported, with the aim to obtain high quality factor in the desiderated spectral range. Non linear second harmonic generation experiments have been performed and the results were compared with bulk GaN sample, highlighting the effect of the microcavity on the non linear optical response of this material.

## 1. INTRODUCTION

III-N compounds exhibit a non-centro symmetric crystalline structure which could provide second order nonlinear optical response ($\chi^{(2)}$)[1] comparable to conventional non linear crystals such as KTP or $LiNbO_3$ (in the visible range). Nevertheless, the large dispersion of bulk nitride material implies too low quadratic nonlinear interaction. Properly designed heterostructures, namely GaN microcavities defined in AlGaN/GaN distributed Bragg reflectors (DBRs), would allow exact phase-matching conditions and simultaneous field localization enhancing nonlinear optical response in this material system. Moreover, the fabrication of 1-dimensional photonic crystals (PhCs) based on these heterostructures can further amplify non linear effects such as second harmonic generation (SHG) and spontaneous parametric down conversion (SPDC), avoiding at the same time the growth of long period structures with high stress level. Several applications like optical communication, multiplexing, frequency conversion or quantum computation would find benefits from such a technology[2].

The limitation for the achievement of a similar objective is twofold: from one side the quality of nitride material and from the other side the reliability of the photonic crystal technology, along with a proper design of the whole system. First of all, the nitride material should exhibit excellent crystalline properties, in order to ensure a highly ordered non centro symmetric structure with the c-axis parallel to the growth direction. Indeed, the crystallinity of AlGaN alloys suffers from the lack of suitable substrates with structural and thermal properties matching those of nitrides. Therefore, GaN and related compounds are generally reported as highly defected materials. It is clear that the disorder induced at a microscale from dislocations and other defects dramatically influences the potential non linear optical response of nitride samples. On the other hand, GaN and AlGaN are robust compounds with large chemical resistance due to strong ionic atomic bond, requiring a suitable development of dry etching techniques (mainly Inductively Coupled Plasma etching), usually based on halogens rather than methane/hydrogen.

In previous studies[3] we have shown that, thanks to a careful study on the growth evolution dynamics in a metal organic chemical vapor deposition (MOCVD) system, the defects in nitride slabs grown on sapphire can be confined into the first 50 nm starting from the interface with the substrate. The strong reduction of the GaN mosaicity achieved in this way is responsible for an enhanced, polarisation dependent second harmonic generation in such films[4] and represents a promising starting point for the realisation of PhCs, aimed to amplify the second order non linear effects in this material.

In this paper we describe the study of SHG enhancement in a GaN microcavity defined by GaN/AlGaN DBRs. Infact, the nonlinear optical response of this material can be further enhanced in a microcavity structure[5] defined by two DBR mirrors, pumped at a fundamental frequency ω and designed for the enhancement of a 2ω mode. GaN/$Al_xGa_{1-x}N$ Distributed Bragg reflectors would allow exact phase-matching conditions and simultaneous field localization[6-7]. The intensity of SH signal has been reported to increase with $L^5$, being L the total length of the multilayered structure, whereas the amplitude of the reflectivity stop band is strongly related to the refractive index difference of the mirror layers. Therefore, a high number of period with a large refractive index difference at the design wavelength is advisable to get a significant enhancement of the transmitted SH signal.

In AlN/GaN systems this implies a high level of tensile stress in the whole structure due to the large lattice mismatch between the two compounds (2.4%). This, in turn, induces a lot of cracks in the DBR when the number of the epitaxial pairs increases, dramatically affecting the DBR reflectance spectrum due to scattering, diffraction and absorption. Therefore, in order to obtain high Q factors without increasing the defect density in the samples, we realised a moderately high Q factor vertical cavity and we planned to further process it according to a 1-D photonic band-gap technology which can further enhance parametric and non-parametric non linear processes. Such a subject is described in details in ref. [8].

The cavity was designed by a software employing the transfer matrix method. The optimal designed sample and all the related test structures were grown by MOCVD, whereas morphological and structural characterisation tools such as atomic force microscopy (AFM), scanning electron microscopy (SEM), high resolution X-ray diffraction (HRXRD) allowed us to assess sample features. Optical measurements in the linear regime were performed by spectrophotometer both in reflectance and transmission configuration and by photoluminescence (PL) with a He-Cd (λ = 325 nm) laser and a Andor CCD for the visible-UV spectral region. The non linear optical response was investigated by measurements of second harmonic generation in non phase-matched non-collinear configuration, as described in ref. [4].

## 2. SAMPLE DESIGN AND GROWTH

### 2.1 Sample design

The microcavity structure described in this work is a GaN-based microcavity grown on sapphire substrate and centred in the near UV spectral range (λ ~400 nm), according to the design provided by transfer matrix method softwares. The AlGaN/GaN microcavity was epitaxially grown on sapphire, by employing a high temperature AlN nucleation layer to allow 2D growth of GaN in a MOCVD system. For this experiment, we chose a fully epitaxial microcavity lay-out. Both top and bottom mirrors consist of 5 period, λ/4 layers of epitaxial GaN and $Al_{0.5}Ga_{0.5}N$ (Δn = 0.4 at 400 nm), whereas the central cavity simply embeds 1 λ GaN (fig.1). A crucial point for the design of the structure is the exact determination of GaN and $Al_{0.5}Ga_{0.5}N$ dispersion law, which is strongly dependent on the measurement method, crystal quality, stress or uncertainty in Al composition for $Al_xGa_{1-x}N$. In this work, we used for GaN, the ellissometric results previously obtained on a bulk sample, whereas for $Al_{0.5}Ga_{0.5}N$ we extrapolated this law from literature values[9-11] again related to a completely relaxed bulk material. However, in both cases, the consequent computation of the optical thickness cannot be considered as completely correct, due to the strong strain field experienced by these materials when they are multistacked in a DBR configuration.

Actually, several studies report how this parameter is also strongly affected by the strain[12]. The solution we adopted in order to get a microcavity centred at the desiderate wavelength of 400 nm was to iteratively adjust the structure parameters thanks to the feedback from SEM, XRD and optical measurements.

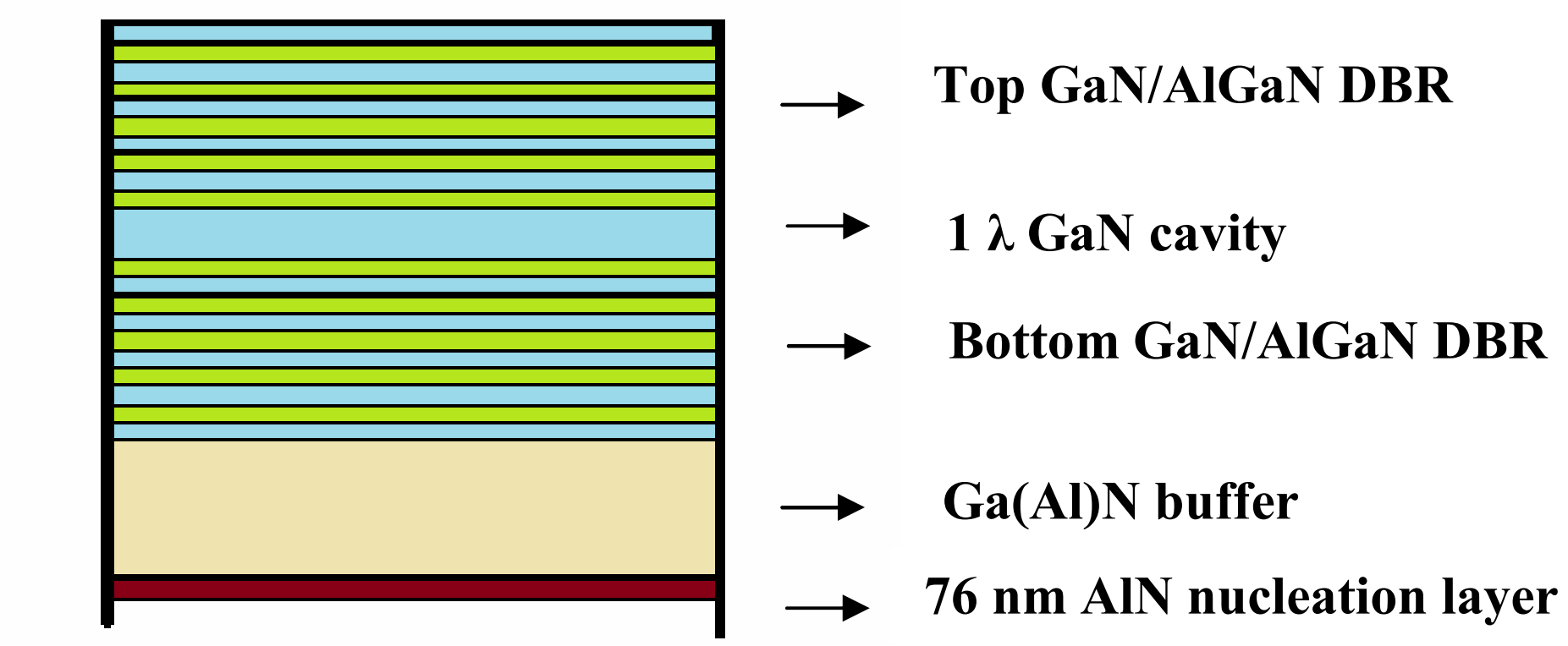

Fig.1: Schematic of the designed GaN/AlGaN microcavity

### 2.2 Growth of microcavity sample by MOCVD

High quality factor cavities, accurately centred at the design wavelength of the pump signal must be obtained for our purposes. To match this requirement several tasks have been fulfilled from the material quality point of view. The first point was the control and optimisation of the AlGaN epilayer growth by MOCVD, where a high compositional uniformity is required to get DBR with the designed reflectivity response. Moreover, the tendency to 3D growth of GaN on AlN due to the high surface energy and the high lattice mismatch between GaN and high Al content AlGaN alloys must also be considered. This in turn produces high surface roughness and high defect density at the GaN/AlGaN interfaces. Therefore, before growing the whole microcavity structures, we performed a detailed investigation on the properties of $Al_xGa_{1-x}N$ layers for different x, ranging from 0.1 to 0.65 and obtained under different growth conditions. In the table below, we report a resume of the different conditions explored.

Table I

| **Al content** | **$NH_3$ pressure (ml)** | **$T_{growth}$ (°C)** | **Reactor pressure (mbar)** | **Growth rate (µm/h)** |
|---|---|---|---|---|
| 0.1-0.65 | 700-2500 | 1150-1180 | 20-50 | 0.4-2.7 |

The following transmission measurements (fig. 2) performed by a spectrophotometer show how the Al composition has been effectively controlled as the related room temperature band edges are in agreement with the expected values for the different Al compositions.

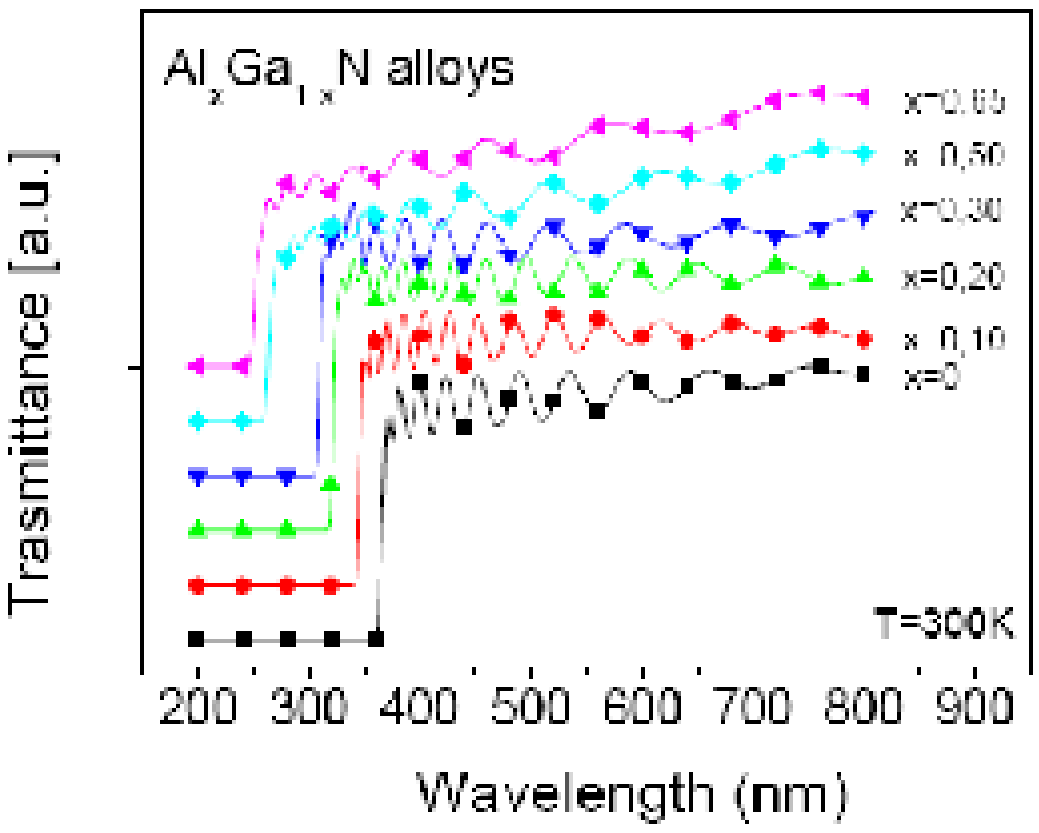

Fig. 2: Transmittance spectra of different Al content ALGaN alloys grown by MOCVD

Particular care was devoted to the optimisation of $Al_xGa_{1-x}N$ epilayers with x = 0.5, which was the DBR designed composition. The effect of the growth parameters listed in table 1 was investigated and we defined a set of optimum growth conditions, compatible with the already stated growth conditions of GaN, which allowed us to obtain the $Al_{0.5}Ga_{0.5}N$ epilayers to be used in the final device. In particular, a low growth rate of 0.45 µm/h was found to favour the homogeneity of Al composition along the layer and to enhance Al ad-atoms mobility, thus allowing a 2-D growth mode with atomic step morphology. Along with a precisely controlled composition, the layer exhibits an extremely low roughness (0.6 nm over a 10µm x 10µm AFM scan range, Fig.3-a) with a morphology typical of a two dimensional planar growth. The good crystalline quality of the layer was confirmed by XRD measurements, with the rocking curve related to the (002) direction of AlGaN as narrow as 57 arcsec. XRD reciprocal space maps (RSMs) were also performed for different symmetric (fig. 3-b) and asymmetric reflections, providing a relaxation value of 42%, indicating that the layer was grown on the AlN nucleation layer with a pseudomorphic strain, without complete relaxation. The tilt angle of the c-axis with respect to growth direction was also extrapolated from X-Ray measurements. Its negligible value (0.02°) allowed us to consider the c-axis of the layer nearly coincident with the optical axis when SHG experiments were performed.

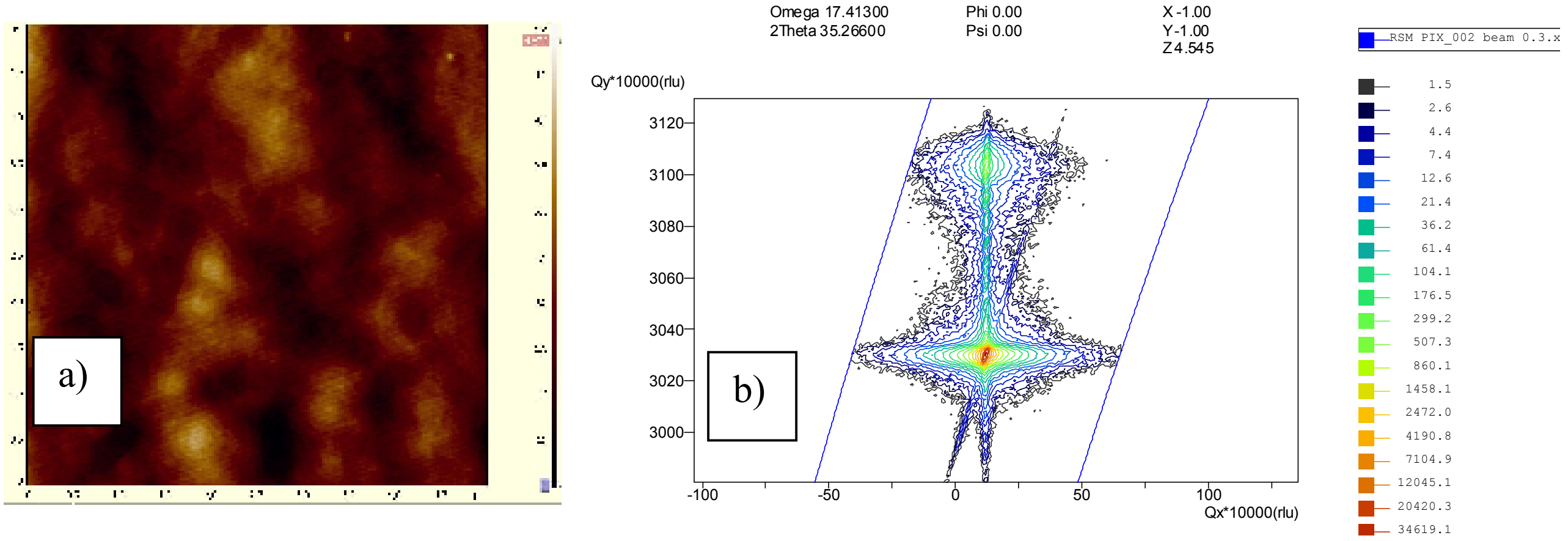

Fig. 3: Morphological characterisation of $Al_{0.5}Ga_{0.5}N$ epilayer by AFM (a) and structural investigation by RSM along the 002 symmetric reflection (b).

Once the optimised growth conditions ensuring good surface roughness were defined, these results were employed for the growth of the whole microcavity sample, according to the design described above. The complete designed structure consists of a 100 nm thick AlN NL, followed by 500 nm of (Al)GaN: these layers serve as a transparent buffer to isolate the microcavity from the substrate. Over there two DBR mirrors consisting of 5 pairs of $Al_{0.5}Ga_{0.5}N$/GaN layers and embedding a λ GaN cavity were grown with thickness related to the design lambda. As already pointed out, the growth of AlGaN/GaN pairs is not straightforward due to the high lattice mismatch among these two materials. For Al content of 50%, the lattice parameter difference is equal to 1.2%, thus meaning that AlGaN grows under a strong tensile strain on the GaN layer underneath. The critical layer thickness of $Al_{0.5}Ga_{0.5}N$ on GaN has been estimated of the order of 40 nm, which is of the order of the design thickness of only one AlGaN layer of our structure. Hence, the formation of cracks propagating along the structure is quite probable to occur. For this reason, before performing the growth of the whole structure some preliminary steps were followed. We first grew some half cavity samples (bottom mirror + GaN cavity) to check the crystalline quality and the optical response of the structures. Considering that the interface quality is one of the most critical aspects in realizing such a kind of structure, we studied the effect of different growth rates of the AlGaN and GaN. Fig. 4 shows the reflectance behaviour of two half cavity samples grown with different growth rate of the AlGaN layers ( 0.45 µm and 0.9 µm). In both samples, it is evident a reflectance dip in the UV region, around 375 nm, whereas the signal drop at the short wavelength side is related to GaN intrinsic absorption. The spectra clearly show an enhancement of the reflectivity peaks when low growth rate is used, even if the central wavelength among the two samples is not altered. The low growth rates increase the mobility of Al adatoms allowing AlGaN growth to occur without the creation of Al rich domains and resulting in a better interface abruptness, responsible of the observed improvement in the reflectivity spectrum. Since the layer thickness were checked by SEM cross section measures, the shift of the central wavelength (with respect to the design one) has been related to a wrong estimation of the $Al_{0.5}Ga_{0.5}N$ refractive index. In fact, in the computation of its optical thickness a refractive index of 2.5 was considered, which is the one of unstrained bulk $Al_{0.5}Ga_{0.5}N$ reported in literature. In this case, the iterative feedback between SEM and reflectance measurements have been useful to centre the final cavity at the designed wavelength.

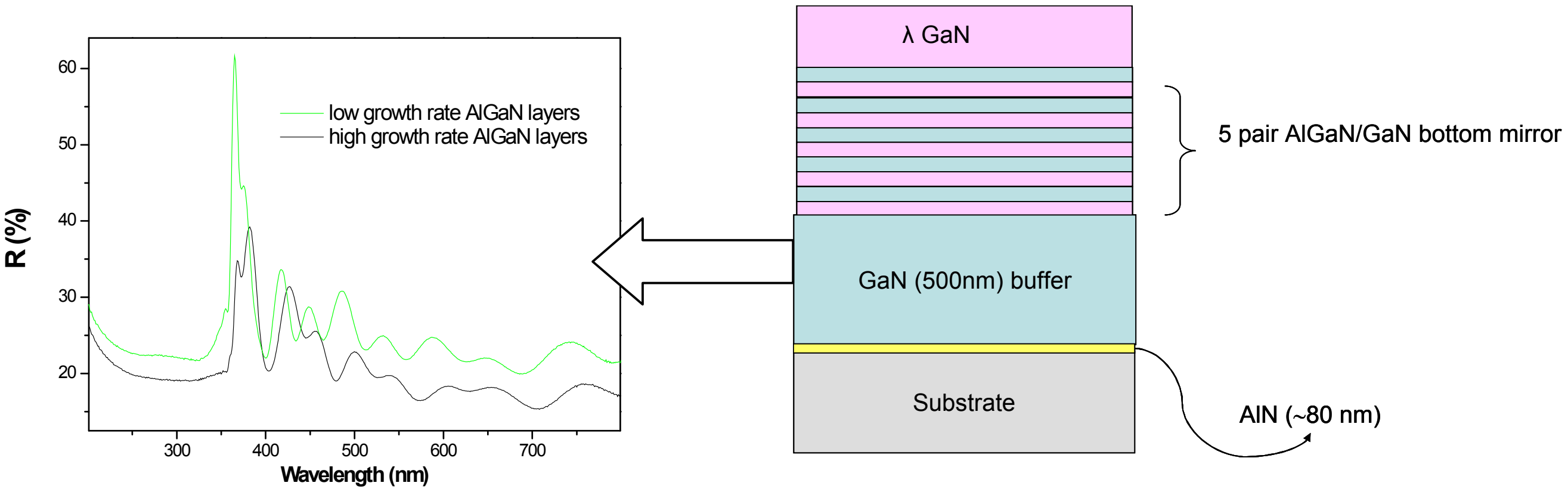

Fig. 4: reflectivity spectra of two half cavities grown by employing two different growth rate for the $Al_{0.5}Ga_{0.5}N$ layers.

The following step was the realisation of whole microcavity samples. We investigated the stress effect on the whole structure by using different buffer compositions (GaN versus $Al_{0.3}Ga_{0.7}N$). As shown in the reflectance spectra of fig. 5-a, this finally lead to an optimised highly symmetric cavity with a 70% reflectance stop band centred at 380 nm with a cavity Q factor ( $\lambda/\Delta\lambda$) equal to 50. The SEM cross section of the optimised structure is shown in fig. 5-b. The image shows very sharp and abrupt interfaces. AFM topography (fig. 5-c) shows the very low roughness of the selected sample, with the step flow morphology typical of a 2D, planar growth.

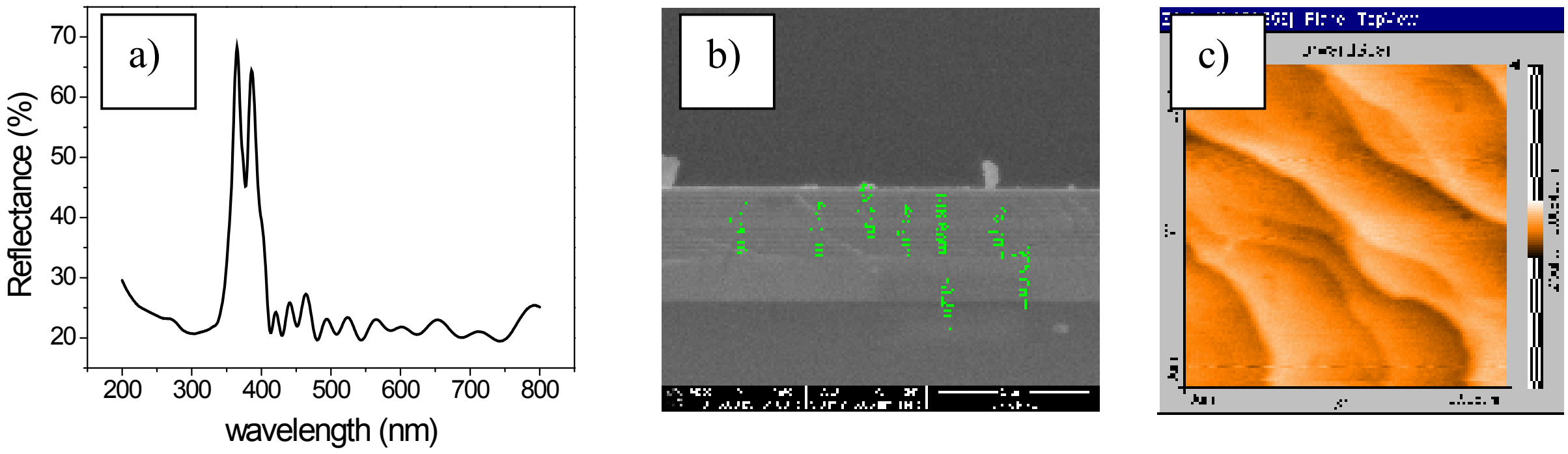

Fig. 5: Reflectance spectrum of the complete microcavity sample (a) and related SEM cross section (b) and AFM scan over 1 μm x 1 μm (c).

## 3. NON LINEAR EXPERIMENTS

We carried out the SHG measurements by means of a noncollinear rotational Maker fringes technique working in the transmission scheme, as reported in Fig. 6. In our experiments the output of a mode-locked Ti:Sapphire laser system tuned at λ=830 nm (76 MHz repetition rate, 130 fs pulse width and average power of 250 mW) was split into two beams of the same intensity using a 50/50 beam splitter. The polarization of both beams can be controlled by two identical rotating half wave plates, which were carefully checked not to give nonlinear contribution since the two collimating lenses, 150 mm focal length, are placed thereafter. The tightly focused beams were sent to intersect in the focus region with an aperture angle of 18° with respect to each other. The sample was placed onto a motorized combined translation and rotation stage which allowed the variation of the sample rotation angle, $\alpha$, with a resolution of 0.05 degrees as well as the z-scan of the sample position within the two beam overlap. Meanwhile, the temporal overlap of the incident pulses was automatically controlled by an external delay line.

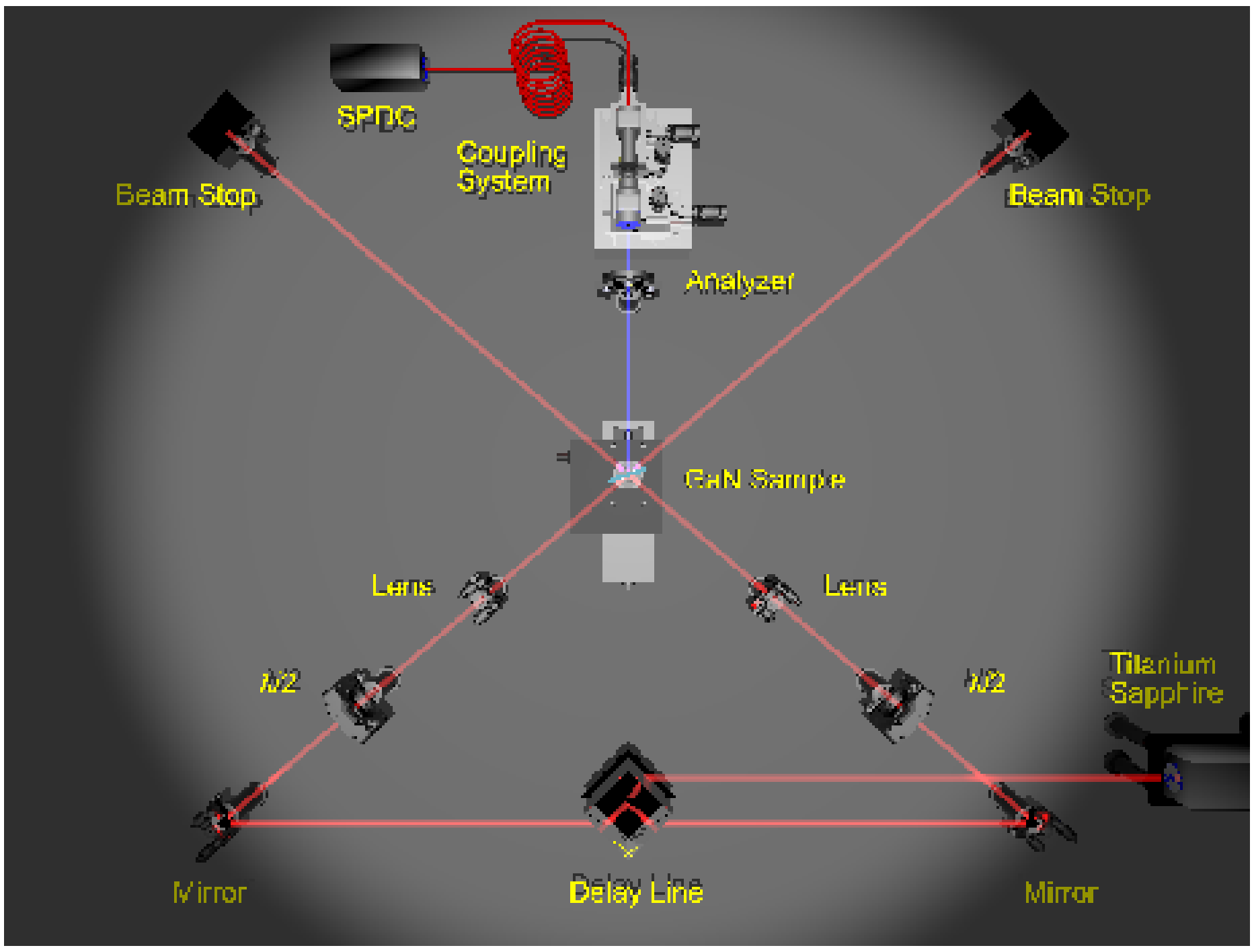

Fig.6 Set up for the SHG experiments at Elsag Datamat, Quantum Optics Lab.

The measurements were carried out at Quantum Optics Lab Elsag Datamat on a reference sample consisting of a 300 nm GaN slab grown on sapphire and on the whole microcavity sample, as shown in fig. 7. GaN bulk sample was excited with a 500 mW pump intensity whereas for the microcavity a power of 250 mW was used, to account for the longer structure. In both cases a strong SH signal is observed. Moreover, for pump beams having same polarization, either -*p* or -s, the generated beam is p-polarized, as well as in the collinear case. In case of crossed polarizations, p-s or s-p, the resulting SH in both GaN and microcavity sample is s polarized, demonstrating that the non linear properties of GaN[4] are preserved in such a configuration. The SH signal profile with the rotation angle shows the higher selectivity of the microcavity sample towards the incident light direction with respect to the GaN bulk sample.

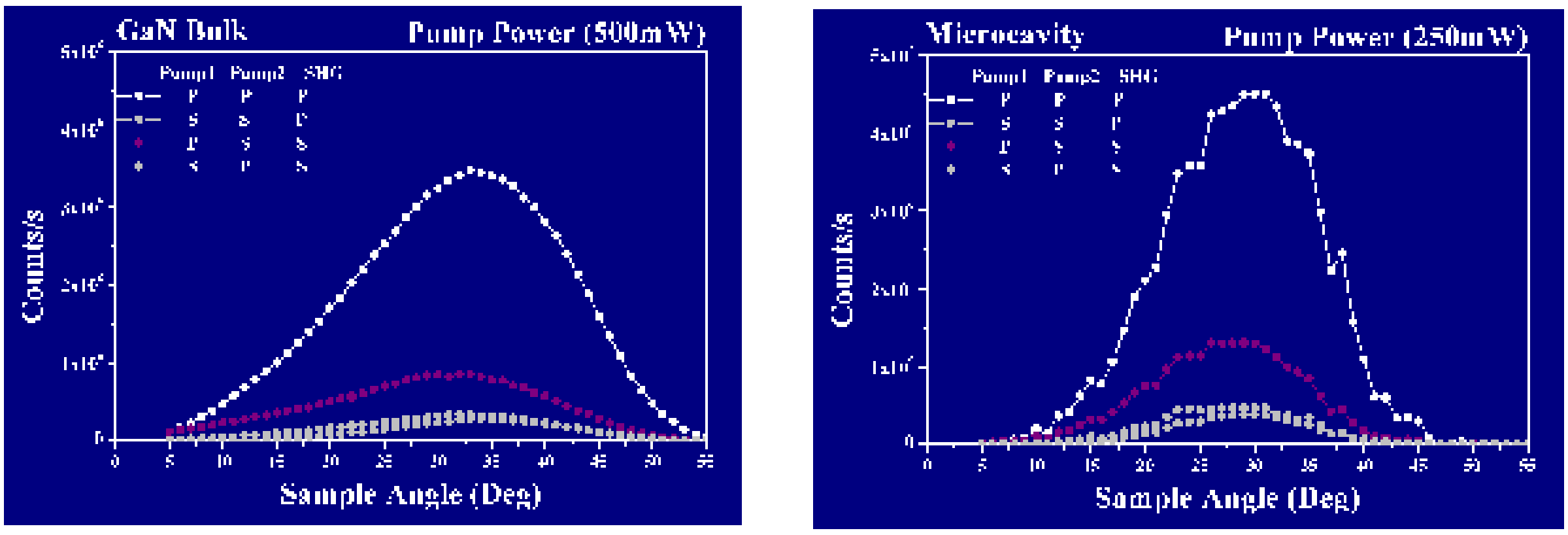

Fig. 7 SH mode intensity as a function of rotation angle for GaN slab and GaN/AlGaN microcavity samples.

## 4. CONCLUSIONS

We have reported on the design, growth and optical characterization of a GaN microcavity defined by multistacked AlGaN/GaN mirrors. The cavity has been designed for the enhancement of second order non linear effects, due to the high non linear response of GaN hexagonal structure. A careful study was performed on the growth evolution of GaN and AlGaN epilayers in order to ensure high material quality, uniform composition and abrupt interfaces. The linear optical characterization shows a cavity with a symmetric reflectance stop band centered in the near UV region (< 400 nm). The cavity was tested in SHG experiments and its behavior was compared with a bulk GaN sample. A bright SH field, strongly dependent on the pump polarization state and with a higher angle selectivity, was generated in the proposed structure, demonstrating the selectivity of the microcavity and the potentialities for applications in non linear field. Further developments can be accomplished by integrating in such a sample innovative photonic crystal geometries, provided that a suitable and reliable technology is developed.

## ACKNOWLEDGEMENTS

This work has been supported by YOUNG-INFM SEED project "Non linear effects in GaN/AlGaN photonic crystals for multiplexing applications" of CNR.